\documentclass[a4paper]{PoS}

\title{European VLBI Network: Present and Future}

\ShortTitle{EVN: Present \& Future}

\author{\speaker{J. Anton Zensus}$^a$
and 
Eduardo Ros$^{abc}$%
\thanks{The authors are grateful to R.W.\ Porcas for careful reading of the manuscript and useful comments and suggestions.}
\\
\llap{$^a$} Max-Planck-Institut f\"ur Radioastronomie\\ 
Auf dem H\"ugel 69, D-53121 Bonn, Germany\\
\llap{$^b$} Observatori Astron\`omic, Universitat de Val\`encia\\ 
C.\ Catedr\'atico Jos\'e Beltr\'an 2, E-46980 Paterna, Val\`encia, Spain\\
\llap{$^c$} Departament d'Astronomia i Astrof\'{\i}sica, Universitat de Val\`encia\\
C.\ Dr. Moliner 50, E-46100 Burjassot, Val\`encia, Spain\\
        E-mail: \email{azensus@mpifr-bonn.mpg.de}, 
\email{ros@mpifr-bonn.mpg.de} 
} 

\abstract{The European VLBI Network\footnote{%
The European VLBI Network is a joint facility of European, Chinese, 
South African and other radio astronomy institutes funded by their 
national research councils.}~is a collaboration of the major
radio astronomical institutes in Europe, Asia, South Africa and Puerto Rico.
Established four decades ago, since then it has constantly improved its
performance in terms made using resolution, data bit-rate and image fidelity with
improvements in performance, and the addition of new stations and
observing capabilities.
The EVN provides open skies access and has over time become a
common-user facility.
In this contribution we discuss the present 
status and perspectives for the array in a continuously
changing environment, especially in the era of ALMA and with the Square 
Kilometre Array \it{ante portas}.
}

\FullConference{12th European VLBI Network Symposium and Users Meeting - EVN 2014,\\
		7-10 October 2014\\
		Cagliari, Italy}

\begin{document}

\section{A very short history of the EVN}

After successful VLBI observations using European telescopes 
in the late 1960s and intra-European observations in the early 1970s, discussions
about a European array date back to 1975.  After several meetings the 
European VLBI Network was formally established in 1980.  A detailed description
of the early times of the network has been published in recent 
conferences \cite{por10,boo12}.  
Initially it was a 4-station network
(Dwingeloo/Westerbork, Effelsberg, Jodrell and Onsala).
From the very beginning a program committee (the EVNPC) 
was established and reviewed observing 
proposals submitted 3 times per year.  A scheduler to coordinate observations
was appointed in 1982, and this task was performed by different individuals
at the MPIfR until Alastair Gunn (Jodrell) took this over  in 2014.  

Since
1980 many stations have been added to the network: 
Medicina (1984),  Wettzell (1995),  Noto (1989),  Shanghai,  
Mets\"ahovi and Cambridge-32m (1990), Yebes-14m and
Urumqi (1994),  Torun-32m  (1996),  Arecibo and
Hartebeesthoek (2001),  Yebes-40m  (2008),    
the KVAZAR network: Svetloe,  Zelenchukskaya and  Badary (2009),    
and recently the KVN: Yonsei,  Ulsan and  Tamna (2014).
The Sardinia-64m and Tianma-65m will start regular participation
in 2015.

Data were initially correlated at the MPIfR processing centre in Bonn.
The Joint Institute for VLBI in Europe, established in 1993,
is currently tasked with data correlation and postprocessing and EVN user support.
A new
correlator was dedicated in 1998,  
and in recent years this has now been replaced by a software
correlator, SFXC.
The EVN has constantly improved its capabilities since foundation, both 
in terms of antenna performance and data bit-rate, 
from the 4\,Mbps provided by the Mk\,II system in 1980 to the plans
for 2 Gbps in 2015 using the Mk\,V system  
with digital backends.

The science addressed by EVN observations  has covered a broad range of 
topics with a high scientific impact.  Several hundred refereed papers 
make use of 
observations from the network;
the most cited works include:

\vspace{2mm}

\noindent
\textbullet~~the properties of compact steep-spectrum  sources using
radio observations at different scales \cite{fan90}\\ 
\textbullet~~a study of a sample of radio galaxies \cite{gio01}\\ 
\textbullet~~examining the dynamics of the compact symmetric objects 0710+439 
and 0108+388 \cite{ows98,ows98b}\\ 
\textbullet~~investigation of methanol masers (a unique EVN capability 
for many years) \cite{min00,beu02}\\ 
\textbullet~~imaging of gravitational lenses such as B0218+35.7 \cite{pat93}\\ 
\textbullet~~imaging of radio jets in galactic objects such as LS\,I\,+61$^\circ$303 
\cite{mas04}. 

\vspace{2mm}

\noindent
Recent publications with the largest number of
citations per month report  astrometric studies of the
Milky Way \cite{rei14,xu13}, 
and feedback in the jet of the radio loud galaxy 4C\,+12.50 
\cite{mor13}.  

\section{Present EVN}

At present the EVN comprises 14 major 
institutes\footnote{See \texttt{http://www.evlbi.org/contact/}.} 
including JIVE.  
Overall EVN  policy is set by the Consortium Board of
Directors. 
%
EVN science topics cover a broad range; they include investigations of
high brightness-temperature objects emitting non-thermal radiation,
 atomic  and molecular processes, synchrotron radiation and
pulsar emission.  The high sensitivity provided by the large collecting
area of many of its elements makes the EVN especially 
suitable for the study of faint  objects such as young radio supernovae.

\paragraph{EVN in context}
In its early days VLBI was confined to centimetre wavelengths
where observing systems made it feasible; shorter wavelengths
suffered from short coherence times and relatively
low performance of receivers, and longer wavelengths were
affected
by the ionosphere and plagued by steadily increasing radio frequency
interference.
Recent technical developments have expanded the parameter space
in radio astronomy. Progress in  
processing data from thousands of dipoles and ``software telescopes'' have
pushed towards longer wavelengths, with pathfinders and
precursors of the Square Kilometre Array (SKA), such as LOFAR.
At the same time, improvements in detector techniques and increasing
bandwidths have triggered the advent of new-generation, short-wavelength 
telescopes, the recently dedicated Atacama Large 
Millimetre/sub-millimetre Array (ALMA) being the most important.
It is
planned that ALMA, as a phased array, will observe as an element
in millimetre VLBI arrays,
notably with the aim of investigating the morphology of the immediate
neighbourhood of the super-massive black holes at the Galactic Centre and
in the nearby galaxy M\,87 -- the so-called Event Horizon 
Telescope \cite{doe10}. 
Additionally, the desire for longer baselines has
pushed VLBI to have elements in Earth orbit, first with  the VSOP
project in the late 1990s \cite{hir98} and currently RadioAstron  \cite{kar13}.
EVN as a ground array is a key element supporting RadioAstron observations.
The use of a large dish such as Effelsberg together with the small
dish onboard \textit{Spektr-R} has the same collecting area as two 
 30-m\,dishes.  The first results reported are spectacular
(see e.g., \cite{kov14} in this conference).

\paragraph{Observing}
The EVN performs observations with disk recording (standard EVN, 
three sessions of three weeks each per year) or in real time (e-VLBI,
10 sessions per year, each of 24h).
Out-of-session scheduling 
has been introduced recently 
in blocks of up to 12 hours of duration
(up to a maximum of 144 hours per year), 
for specific purposes which
justify observations outside of regular sessions.
Observations are possible at 92, 49, 30, 21, 18, 13, 6, 
5, 3.6, 1.3, and 0.7\,cm wavelength.  
Joint observations with the Very Long Baseline Array (VLBA),
('Global' proposals) can also include  the Green Bank 
Telescope and the phased Jansky Very Large Array.
Global proposals can currently use up to 1\,Gbps data bit rate.  
Joint observationswith the RadioAstron 
project  are possible as well.  Following the Korean VLBI Network joining 
the EVN as an  associate member in 2014, it is planned that some joint 
time with the Australian Long Baseline Array will be available, starting in 
2015, creating a real global array.  The details and updates of 
the EVN performance are announced in every call for 
proposals\footnote{See \texttt{http://www.evlbi.org/proposals/call.txt}. }.  
The status of the 
telescopes, receiver availability, observing modes, etc., is 
maintained by JIVE in the EVN status 
tables\footnote{See \texttt{http://www.evlbi.org/user\textunderscore guide/EVNstatus.txt}. }.  
For observations at 3.5\,mm, 
astronomers can use the
Global Millimetre VLBI Array (GMVA),
operated jointly by the MPIfR, IRAM, Onsala and 
NRAO\footnote{See \texttt{http://www3.mpifr-bonn.mpg.de/div/vlbi/globalmm/}. }.
\subsection{Recent highlights}

\paragraph{Technical}
As  mentioned above the KVN recently joined the EVN.  
Its 3 antennas (with separations up to 480\,km)
at the easternmost edge of the network enhance 
remarkably the $(u,v)$ coverage.  An interesting feature of
these telescopes is their novel high-frequency capabilities,
since {\it simultaneous} observations are possible at 
13, 7, 3.5, and 2\,mm wavelength; 
13 and 7\,mm are offered by the EVN for joint observations.  

RadioAstron has performed joint observations with the EVN since early 2012,
combining the high resolution provided by space VLBI with the high sensitivity
of the EVN (see above).

Recently a new correlator developed at JIVE (UniBoard project, supported
by RadioNet3, see \cite{szo10}) has successfully demonstrated
4\,Gbps operation.
When brought into operation the correlator will support 
32 stations with 64\,MHz bandwidth, integration
times of 0.022\,s to 1\,s, and a frequency resolution up to 15.625\,kHz.

In September 2014 the  radome of the 20m-Onsala telescope was renewed.
The top cap of 50 panels was replaced in one piece, and the remaining
570 elements were changed one by one\footnote{See 
\texttt{http://goo.gl/Ca0dJH}.
}.

As mentioned above the Tianma 65-m telescope will be available for EVN observations in 2015. 
It will operate with
adaptive optics in all the bands offered by the EVN from 21\,cm to 7\,mm.
First fringes with the EVN were obtained in March 2014 \cite{she14}.
It can also provide a very short baseline together with the Seshan 25-m telescope.

\paragraph{EU Support}
The EVN is an excellent example of international cooperation in science.
This collaborative aspect has been boosted over 
the last twenty years by the support of 
the European Commission
via different funding instruments.
Following from support 
inder the 3$^\mathrm{rd}$ Framework Program (FP3) with 
\textsl{The European VLBI network of radio 
telescopes}\footnote{Ref.\ CHGE920011,
programme FP3-HCM.}, 
it  continued
in FP4 with
\textsl{Enhancing the European VLBI Network of Radio 
Telescopes}\footnote{Code FMGE980101, programme FP4-TMR.} 
and
\textsl{Access to the EVN of radio telescopes}\footnote{Code FMGE950012, 
programme FP4-TMR, subprogramme 0201 - Access for researchers, 
funding scheme 0201 - Access for researchers.},
%
in FP5 with 
the projects
\textsl{EVN-ACCESS}\footnote{\textsl{European vlbi network via the joint institute for vlbi in europe}, Ref.\ HPRI-CT-1999-00045, programme FP5-HUMAN POTENTIAL.}
%
and
\textsl{FARADAY}\footnote{\textsl{Focal-plane arrays for radio astronomy; design, access and yield}, 
Ref.\ HPRI-CT-2001-50031, 
programme FP5-HUMAN POTENTIAL.},
in FP6 with
\textsl{RadioNet}\footnote{\textsl{RadioNet: Advanced Radio Astronomy in Europe}, 
Ref.\ 505818, 
programme FP6-INFRASTRUCTURES, 
subprogramme INFRASTR-2.1 - Integrating activities combining cooperation 
networks with transnational access and research projects.}
and
\textsl{EXPReS}\footnote{\textsl{EXPReS: a production astronomy e-VLBI infrastructure},
Ref.\ 026642, 
programme FP6-IST, 
funding scheme I3 - Research Infrastructure-Integrated Infrastructure 
Initiative.},
%
and
in FP7 with
\textsl{NEXPReS}\footnote{\textsl{NEXPReS- Novel 
EXplorations Pushing Robust e-VLBI Services}, 
Ref.\ 261525, 
programme FP7-INFRASTRUCTURES, 
subprogramme INFRA-2010-1.2.3 - Virtual Research Communities.},
%
\textsl{RadioNet-FP7}\footnote{\textsl{Advanced Radio Astronomy in Europe},
Ref.\ 227290,
programme FP7-INFRASTRUCTURES, 
subprogramme INFRA-2008-1.1.1 - Bottom-up approach: 
Integrating Activities in all scientific and technological fields.}
and
currently
\textsl{RadioNet3}\footnote{\textsl{Advanced Radio Astronomy in Europe},
Ref.\ 283393,
programme FP7-INFRASTRUCTURES, 
subprogramme INFRA-2011-1.1.21. - Research Infrastructures for advanced radio astronomy.}.
%

Additional funding for EVN research activities was provided by programs such as
the cooperation with the former Soviet Union by
\textsl{The nature and origin of the most compact cosmic 
radio sources known in the Universe}\footnote{Ref. INTAS-94-4010.},
geodesy-related projects such as 
\textsl{RADIO-INTERFEROMETRY}\footnote{\textsl{Measurement of vertical 
crustal motion in Europe by VLBI}, Ref.\ FMRX960071, programme FP4-TMR,
subprogramme 1.4.1.-3.1S4 - Environment and Geosciences.},
and
research training networks such as
\textit{CERES}\footnote{\textsl{The universe at high redshift and the physics of active galaxies from multi-wave length studies of compact radio sources - consortium for European research on extragalactic surveys}, 
Ref.\ FMRX960034,
programme FP4-TMR,
subprogramme 1.4.1.-3.1S7 - Physics.},
\textit{ANGLES}\footnote{\textsl{Astrophysics Network for Galaxy Lensing Studies}, Ref.\ 505183, programme FP6-MOBILITY, subprogramme 
MOBILITY-1.1 - Marie Curie Research Training Networks (RTN).}
and 
\textsl{ESTRELA}\footnote{\textsl{Early stage training site for European long-wavelength Astronomy}, 
Ref.\ 19669,
programme FP6-MOBILITY,
subprogramme MOBILITY-1.2 - Marie Curie Host Fellowships - 
Early stage research training (EST).}.

The present I3 project \textsl{RadioNet3}
includes a transnational access
programme (which supports EVN observations and data analysis), 
networking activities (which, amongst other things, support
this conference), and joint reseach activities to support research and 
development at the radio astronomical facilities in Europe, including
JIVE and several EVN radio telescopes.   
A complete description of the
project and its goals is provided in its webpage\footnote{See
\texttt{http://www.radionet-eu.org/}.}.  \textsl{RadioNet3} also
contributes to the implementation of the strategic plan for
European radio astronomy (\textsl{AstroNet}\footnote{See 
\texttt{http://www.astronet-eu.org/}.}) by building a 
sustainable radio astronomical research community.

%

\paragraph{Scientific}
As reported above, the EVN has produced hundreds of scientific publications
in several areas; highlights are shown regularly on the EVN webpages.
Its high fidelity imaging is especially useful for global 
experiments and for observations of complex jet structures, e.g., in the
study of helical features in 0836+710 \cite{per12}.  Methanol masers
can be probed at 5\,cm wavelength, and these astrometric results complement
astrometric studies performed at 1\,cm with water masers, see \cite{ryg12}.
The high sensitivity shows its full power in the study of 
ultraluminous X-ray sources \cite{mez13}, in setting upper limits
to the emission from radio supernovae, e.g., SN\,2014 \cite{per14}, or
in the imaging of the Crab nebula \cite{lob11}.  
Synergies with observations from telescopes at the very highest frequencies
are possible, as in the case of the blazar IC\,310 in the 
Perseus cluster, studied jointly by MAGIC and the EVN \cite{ale14}.

\section{The future}

\paragraph{JIV-ERIC} The future has now become the present, since the European Commission decided
in the course of writing this contribution to 
allow JIVE to become a European Research Infrastructure Consortium, 
initially with four  member countries (The Netherlands, United Kingdom, Sweden and France).
In addition, research councils and institutes in other countries --
Italy, Spain, South Africa, Germany and China -- will contribute
to JIVE as well.  The official dedication  of the JIV-ERIC is planned for 
April 2015.

\paragraph{Expanding the network}  Six dishes were recently  added to the network (KVAZAR with 
3$\times$32\,m in 2009 and KVN with 3$\times$21\,m in 2014).
The telescopes in Sardinia (64-m) and Tianma (65-m) are a reality
and will increasingly participate in EVN observations in 2015.
Further in the future are the planned 110-m telescope in Qitai, near Urumqi, and 
the FAST 500-m telescope in China, which could join EVN observations,
and also a 70-m class telescope in Poland (RT90) to the north of Toru\'n. 
Furthermore, a beam-formed MeerKAT telescope would be an 
excellent addition to the network to the South, with a short
baseline to Hartebeesthoek and with high sensitivity.  Plans for
an African VLBI array are also being developed, and this would
yield an unprecendented coverage at centimetre wavelengths, to 
boost radio astronomy in the next years, complementing the Square
Kilometre Array.


\paragraph{New frontiers}
In the future the EVN will continue being a key instrument
in radio astronomy, complementing multi-messenger campaigns, 
in the study of transients, enhancing its image fidelity and
astrometric precision, complementing millimetre- and space-VLBI, and
as a match in the radio for astrometric measurements in the
era of \textsl{GAIA}.  Even when other facilities are threatened
with funding cuts, one of the strengths of the EVN is its international
nature, making a sustainable and flexible facility for many years
in the future.



\renewcommand{\baselinestretch}{0.9}



\small
\begin{thebibliography}{99}
\bibitem{por10} Porcas, R.W., \pos{PoS(10th EVN Symposium)011} (2010)
\bibitem{boo12} Booth, R.S., \pos{PoS(RTS2012)005} (2012)
\bibitem{fan90} Fanti, R., et al., A\&A 231, 333-346 (1990) 
\bibitem{gio01} Giovannini, G., et al., ApJ 552, 508--526 (2001) 
\bibitem{ows98} Owsianki, I. \& Conway, J.E., A\&A 337, 69--79 (1998) 
\bibitem{ows98b} Owsianik, I., Conway, J.E., \& Polatidis, A.G., A\&A 336, L37 (1998b) 
\bibitem{min00} Minier, V., Booth, R.S., \& Conway, J.E., A\&A 362, 1093--1108 (2000) 
\bibitem{beu02} Beuther, H., et al., A\&A 390, 289--298 (2002) 
\bibitem{pat93} Patnaik, A.R., et al., MNRAS 261, 435--444 (1993) 
\bibitem{mas04} Massi, M., et al., A\&A 414, L1-L4 (2004) 
\bibitem{rei14} Reid, M.J., et al., ApJ 738, 130 (2014) 
\bibitem{xu13} Xu, Y., et al., ApJ 769, 15 (2013) 
\bibitem{mor13} Morganti, R., et al., Science 341, 1082 (2013) 
\bibitem{doe10} Doeleman, S.D., \pos{PoS(10th EVN Symposium)053} (2010) 
\bibitem{hir98} Hirabayashi, H., et al., Science 281, 1825 (1998) 
\bibitem{kar13} Kardashev, N.S., et al., Astron.\ Rep.\ 57, 153--194 (2013) 
\bibitem{kov14} Kovalev, Y.Y., et al., \pos{PoS(EVN 2014)020} (2014) 
\bibitem{szo10} Szomoru, A., \pos{PoS(10th EVN Symposium)098} (2010) 
\bibitem{she14} Shen, Z.Q., in 3rd China-U.S. Workshop on Radio Astronomy Science and Technology, see \verb+http://goo.gl/5N7M6u+ (2014) %
\bibitem{per12} Perucho, M., et al., ApJ 749, 55 (2012) 
\bibitem{ryg12} Rygl, K.L.J., et al., A\&A 539, A79 (2012) 
\bibitem{mez13} Mezcua, M., et al., MNRAS 436, 1546--1554 (2013) 
\bibitem{per14} P\'erez-Torres, M.A., et al., ApJ 792, 38 (2014) 
\bibitem{lob11} Lobanov, A.P., et al., A\&A 533, A10 (2011) 
\bibitem{ale14} Aleksic, J., et al., Science 346, 1080--1084 (2014) 
\end{thebibliography}
\end{document}